\documentclass[10pt,twocolumn,showpacs,amsmath,amssymb,aps,prl,longbibliography,superscriptaddress]{revtex4-1}
\usepackage{times}
\usepackage[utf8]{inputenc}
\usepackage{newunicodechar}
\newunicodechar{，}{,}
\usepackage{color}
\usepackage{array}
\usepackage{verbatim}
\usepackage{multirow}
\usepackage{amsmath}
\usepackage{amssymb}
\usepackage{ytableau}
\usepackage{tabularx}
\usepackage{subfigure}
\usepackage{graphicx}
\usepackage{esint}
\usepackage{booktabs}
\usepackage[unicode=true,
bookmarks=true,bookmarksnumbered=true,bookmarksopen=false,
breaklinks=true,pdfborder={0 0 0},backref=false,colorlinks=true]
{hyperref}
\usepackage{enumitem}
\usepackage{cleveref}
\usepackage{xcolor}
\usepackage[bb=boondox]{mathalfa}
\usepackage[title]{appendix}
\usepackage{makecell}
\usepackage[T1]{fontenc}
\usepackage{amsmath}
\usepackage{slashed}
\usepackage{ulem}

\hypersetup{
	linkcolor=red,          
	citecolor=blue,        
	filecolor=magenta,      
	urlcolor=magenta           
}

\usepackage{cleveref}
\crefname{appendix}{Appendix}{Appendices}
\crefname{equation}{Eq.}{Eqs.}
\crefname{figure}{Fig.}{Figs.}
\crefname{table}{Table}{Tables}
\crefname{section}{Sec.}{Secs.}

\renewcommand{\paragraph}[1]{\vspace{0.2cm}{\bf \textit{#1}}}

\definecolor{Gray}{gray}{0.85}
\newcolumntype{a}{>{\columncolor{Gray}}c}

\usepackage{simpler-wick}
\allowdisplaybreaks[1] 

\begin{document}
\title{Incompressible quantum Hall liquid on the four-dimensional sphere }
\author{Junwen Zhao }
\thanks{These two authors contributed equally to this work} 
\affiliation{Department of Physics,	Fudan University, Shanghai 200433,China}
\affiliation{New Cornerstone Science Laboratory, Department of Physics, School of Science}
\affiliation{Department of Physics, School of Science, Westlake University, Hangzhou 310024, Zhejiang, China}
\author{Xue Meng }
\thanks{These two authors contributed equally to this work} 
\affiliation{Department of Physics,	Fudan University, Shanghai 200433,China}
\affiliation{Department of Physics, School of Science, Westlake University, Hangzhou 310024, Zhejiang, China}
\author{ Wei Zhu }
\email{weizhu@westlake.edu.cn}
\affiliation{Department of Physics, School of Science, Westlake University, Hangzhou 310024, Zhejiang, China}

\author{Congjun Wu }
\email{wucongjun@westlake.edu.cn}
	\affiliation{New Cornerstone Science Laboratory, Department of Physics, School of Science}
	\affiliation{Department of Physics, School of Science, Westlake University, Hangzhou 310024, Zhejiang, China}
	\affiliation{Institute for Theoretical Sciences, Westlake University, Hangzhou 310024, Zhejiang, China}
	\affiliation{Key Laboratory for Quantum Materials of Zhejiang Province,}
	\affiliation{Institute of Natural Sciences, Westlake Institute for Advanced Study, Hangzhou 310024, Zhejiang, China }
	\date{\today}
	\begin{abstract}
The quantum Hall effect (QHE) is a cornerstone of topological physics, inspiring extensive explorations of its high-dimensional generalizations such as experimental realizations in synthetic systems including cold atoms, photonic lattices and metamaterials. 
However, the many-body effect in the 
higher dimensional QHE system remains poorly understood. 
We explore this problem by formulating the microscopic wavefunctions on a four dimensional sphere  inspired by Laughlin’s seminal work. 
Employing a generalized pseudo-potential framework, we derive an exact microscopic Hamiltonian consisting of two-body projection operators that annihilate the microscopic wavefunctions. Diagonalizations on finite system sizes show that the quasi-hole states remain zero energy while the quasi-particle states exhibit a finite gap, in consistency with an incompressible state. 
Furthermore, the pair distribution is calculated to substantiate the liquid-like nature of the wavefunction. 
Our work provides a preliminary understanding to the fractional quantum Hall states in high dimensions. 
\end{abstract}
\maketitle
	
    The discovery of the integer quantum Hall effect (IQHE) \cite{Klitzing1980} marked a pivotal milestone in condensed matter physics
    research, opening up an era for studying topological phenomena.
    In IQHE systems, two-dimensional  electrons subjected to a magnetic field form Landau levels, and the quantization of Hall conductance is intrinsically tied to the Landau level wavefunction topologies \cite{Thouless1982}. 
	Moreover, many-body properties become 
    research focus following the discovery of the fractional quantum Hall effect (FQHE) \cite{Stormer1982}. 
    The groundbreaking insight comes from the celebrated Laughlin wavefunction \cite{Laughling1983}, which captures the essence of the FQHE, including fractionally quantized Hall conductance and fractionally charged excitations. 
    Various generalizations of the Laughlin wavefunction have been proposed, including the Read-Moore state \cite{Moore1991} and the Haldane-Rezayi state \cite{Haldane1988}, offering new perspectives on the FQHE.
	

%
    
	Various investigations have explored the extension of Landau levels to higher-dimensional systems \cite{zhang_2001_a,Nair2003,Bernevig2002-o,JELLAL2005554,w2012,Wu2013,w2013,DaweiWang2025}.
    Most notably, Landau levels on the four-dimensional (4D) sphere ($S^4$) have been constructed \cite{zhang_2001_a} by introducing a Yang monopole located at the origin \cite{Yang197812}, analogous to Haldane's construction of Landau levels on $S^2$ with a Dirac monopole \cite{Haldane1983}.
    The 4D quantum Hall effect (4D QHE) has been studied in the context of noncommutative geometry
    \cite{ YixinChen2000,susskind2001,Hou2002,
   zhang2002,Hasebe2010,hesebe2018, hesebe2020, hesebe2022,hasebe2023}.
   Experimental advances have also demonstrated the realization of the 4D QHE in cold atom systems \cite{science2024, lohse_2018_exploring}, optical lattices \cite{zilberberg_2018_photonic},  acoustic lattices  \cite{GuancongMaprx2021,Chen2021,Xu2022,Chen2023NSR,
    Chenyanfeng2024}, and electric circuits\cite{Price2020,BaiLeZhang2020,Yu2020NSR4DSpinless,Zhang2020PRB4DHexadecapole}. 
	Furthermore, the Yang monopole has been simulated in cold atom experiments and metamaterials
    \cite{sugawa_2018_second, shuangzhang2023,Zhangshuangprl2023}.
    Nevertheless, the many-body physics of the 4D FQHE remains poorly understood, constituting an exceptionally difficult and largely unexplored frontier 
  for investigations 
    via the synthetic dimension \cite{phys2015,Yang_2024,Price2019,QuanShengWu2019, Shuangzhang2021, Shuangzhang2021nature, Ewaza2021, Segev2021,Ryderbergsynthetic2022,Thoulesspumping2024}  and the hyperbolic lattice \cite{Kollar2019_HyperbolicLattices,Hyperbolic——band,Hyperbolic-PNAS,Hyperbolic-irreps,XiangdongZhang2023_HyperbolicC2,Xiangdong2025}  in future experiments.

	In this article, we explore the many-body properties based on $S^4$ 
   augmented by a
    Yang monopole background.
    Both the determinant-type and 
    Jastrow-type Laughlin wavefunctions are constructed, which are the ground states of Hamiltonians with suitably designed  pseudo-potentials. 
    Exact diagonalizations are performed 
    to 
    the corresponding Hamiltonians, revealing signatures of an incompressible quantum liquid state.
    These states provide further insights for studying high dimensional interacting topological phases. 
 
We begin with the following single-particle Hamiltonian
 defined on an $S^4$ sphere \cite{Yang197812,zhang_2001_a},
	\begin{equation}
	H=\frac{(P_{\mu}+  A_{\mu})^2}{2M} , 
	\label{eq1}
	\end{equation}
	where the radius is $R$;
    $A_{\mu}=-\frac{1}{2R(R+x_5)} \eta_{\mu\nu}^{a} I_a x_{\nu}$ is the SU(2) gauge field of a Yang monopole, with the t'Hooft symbol $\eta_{\mu\nu}^{a} (\mu ,\nu =1,2,3,4;a=1,2,3)$ \cite{Hooft1976}; $I_a$ are the SU(2) generators with 
 $I$ and $I(I+1)$ the values of the isospin and Casimir, respectively. 
    The single-particle wavefunctions of Eq.(\ref{eq1}) are organized into the SO(5) irreducible representation (IRREP)  \footnote{We emphasize that the projective representations of group SO(5) are employed as the single-particle bases on $S^4$ to construct many-body wavefunctions. The lowest Landau level wave functions belong to the representation of $(2I,0)_{\text{\tiny SO(5)}}$, which is projective when I is a half-integer, or, equivalently, a spinor representation of Sp(4) or Spin(5).  Sp(4) is the double covering group of SO (5) with the same Lie algebra. The former possesses spinor representations, but the latter does not. } $(N+2I,N)_{\text{\tiny {SO(5)} }}$\cite{Yang1978,Yang197812,zhang_2001_a}, where $N\ge 0$ is the Landau level index. 
	The eigen-energy is given by
         $E_N=\frac{\hbar^2}{2MR^2} (\mathcal{C}(N+2I,N)-2I(I+1))$, 
where $\mathcal{C}(p,q)=\frac{p^2+q^2}{2}+2p+q$ is the eigenvalue of the SO(5) Casimir.
The lowest Landau level (LLL) states form the IRREP of $(2I,0)_{\text{\tiny {SO(5)} }} $, whose degeneracy  is given by
    \begin{align}
        d(2I,0)=\frac{1}{3!}(2I+1)(2I+2)(2I+3).
\label{eq:degneracy}
    \end{align}

A brief introduction to the 
$ \text{so}\text{\small{(5)}}$ 
algebra in the presence of a Yang monopole is presented in End Matter. We decompose the LLL states into the IRREPs of the subgroup SO(4), whose six generators can be constructed with the  t'Hooft symbol
\footnote{ The $\hat{J}_a,\hat{K}_a$ satisfy commutation that 
    $[\hat{J}_a,\hat{J}_b]=\text{i}\varepsilon_{abc}\hat{J}_c,
    [\hat{K}_a,\hat{K}_b]=\text{i}\varepsilon_{abc}\hat{K}_c,
    [\hat{J}_a,\hat{K}_b]=0 $, see more details in S.M.~\cite{supplement}
    } 

\begin{eqnarray}
\hat{J}_a=\frac{1}{4}\eta^a_{\mu\nu}L_{\mu\nu}, \ \ \,
\hat{K}_a=\frac{1}{4}\bar \eta_{\mu\nu}^{a} L_{\mu\nu}. 
\label{eq:so4_gen}
\end{eqnarray}
    Applying the branching rule to the LLL states, we obtain $(2I,0)_{\text{\tiny {SO(5)} }}=\bigoplus\limits_{j+k=I}(j,k)_{\text{\tiny {SO(4)} }}$.
    Since $\hat{J}_a$ and $\hat{K}_a$ form two mutually commuting su\text{\small(2)} algebras, $(j,k)_{\text{\tiny {SO(4)} }}$ denotes the direct product of two SU(2) representations characterized by quantum numbers $j$ and $k$, respectively. 
	

Next we explain how to express the LLL wavefunctions  by the
homogeneous polynomials of the 4-component fundamental SO(5) spinors $\psi_\alpha$ with $\alpha=1\sim 4$.
Through the 2nd Hopf map,
one define a unit vector $x_a = \psi^\dagger_\alpha \Gamma^a_{\alpha\beta} \psi_\beta ,(a=1,\dots,5)$ on $S^4$.
The $4\times 4$ Clifford algebra matrices $\Gamma^a$ mutually anticommute and their explicit representation is presented in Sec.~II of S.M. \cite{supplement}.
Conversely, each $\psi_\alpha$ can be expressed through $x_a$ and a 2-component spinor $u=(u_1, u_2)^T$ with the isospin coordinates on $S^2$ defined via the 1st Hopf map ${n}_i=u^\dagger {\sigma}_i u $.
The LLL wavefunctions can be organized into a $(2I+1)$-component spinor denoted by $f_{I_3}(x_a)|I I_3\rangle$,
with each component being an eigenstate of $I_3$ satisfying $I\geq I_3 \geq -I$.
For compactness, $|II_3\rangle$ is represented on $S^2$  by
$u_1^{I+I_3} u_2^{I-I_3}/\sqrt{(I+I_3)!(I-I_3)!}$.
Then the LLL wavefunctions  are expressed as
\begin{eqnarray}
f_{j,m_1;k,m_2}^{(2I,0)} (x_a,n_i)=
{\cal N}
\psi_1^{j+m_1}\psi_2^{j-m_1}\psi_3^{k+m_2} \psi_4^{k-m_2}, 
\label{eq:LLL}
\end{eqnarray}
where 
${\cal N}=1/\sqrt{{(j+m_1)!(j-m_1)!(k+m_2)!(k-m_2)!}}$ is the normalization factor .



For later convenience in exploring many-body wavefunctions, we first construct the two-body states under the SO(5) rotational symmetry on $S^4$, which is significantly more complex than the $S^2$ case under the SU(2) symmetry\cite{Haldane1983}. 
The product of two single-particle LLL states can be decomposed into different SO(5) channels.
According to the decomposition of the direct product of SO(5) IRREPs
\footnote{As detailed in Sec.~IX of
S.M.~\cite{supplement}, 
the CG series of the SO(5)  decomposition are derived by leveraging the orthogonality of group characters.
},
we have
\begin{eqnarray}
(2I,0)_{\text{\tiny {SO(5)} }} 
\otimes (2I,0)_{\text{\tiny {SO(5)} }} = \bigoplus\limits_{n_1,n_2} (n_1+n_2,n_1-n_2)_{\text{\tiny {SO(5)} }},
\label{eq:decompose}
\end{eqnarray}
where $0\le n_1\le 2I$,
and $0\le n_2\le n_1$.
The simplest set of the two-body states lie in the channel of $(4I-2n,0)_{\text{\tiny {SO(5)} }}$ with $0\le n\le 2I$.
An SO(5) invariant is constructed as $S(x_a,n_i;x^\prime_a,n_i^{\prime})=\psi_{\alpha}(x_a,n_i) \mathcal{R}_{\alpha\beta} \psi_{\beta}(x^\prime_a,n_i^{\prime}) $ with 
$\mathcal{R}$ being the charge conjugation matrix for the SO(5) 
fundamental spinor, satisfying 
$\mathcal{R}^2=-1$, $\mathcal{R}^{\mathbf{T}}=\mathcal{R}^{-1}=\mathcal{R}^{\dagger}=-\mathcal{R}$ \footnote{The explicit definition of $\mathcal{R}$ is shown in Sec.~III of S.M.~\cite{supplement}. The $\mathcal{R}$ is defined as $\mathcal{R}=-\text{i} \Gamma_3 \Gamma_1$}.
With the help of $\mathcal{R}$, we obtain the two-body states in the form
of
\begin{eqnarray}
&\Phi_{J,M_1;K,M_2}^{(4I-2n,0)}(
x_a,n_i;
x^\prime_a,n_i^{\prime})
=  \mathcal{N}_{J,M_1;K,M_2}^{(4I-2n,0)}   
S^n \nonumber \\
&\times
f_{J,M_1;K,M_2}^{4I-2n}(x_a,n_i;x^\prime_a,n_i^{\prime}), 
\label{eq:2-body1}
\end{eqnarray}
where $\mathcal{N}_{J,M_1;K,M_2}^{(4I-2n,0)}$ is the normalization coefficient.
The center-of-mass  part $f_{J,M_1;K,M_2}^{4I-2n}$ is a symmetric homogeneous polynomial with the degree  $4I-2n$ defined as
\begin{eqnarray}
f_{J,M_1;K,M_2}^{4I-2n}(x_a,n_i;
x^\prime_a,n_i^{\prime})
=C_{j,m_1;j^{\prime},m_1^{\prime}}^{J,M_1} C_{k,m_2;k^{\prime},m_2^{\prime}}^{K,M_2} 
\nonumber\\
\times 
f^{(2I-n,0)}_{j,m_1;k,m_2}(x_a,n_i)
f^{(2I-n,0)}_{j^{\prime},m_1^{\prime};k^{\prime},m_2^{\prime}}(x^\prime_a,n_i^{\prime}),
\end{eqnarray}
where $C_{j,m_1;j^{\prime},m_1^{\prime}}^{J,M_1}$ is the  SU(2) group Celebsh-Gordon (CG) coefficient.
The statistics of the two-body wavefunciton is captured by the power of the relative part $S^n$: Odd and even values of $n$ correspond to the fermionic and bosonic statistics, respectively.

We move forward to the two-body states in more complex channels with $n_1>n_2$.
Take the case of
$n_1-n_2=1$ as an example
which involves the regular SO(5) harmonics in the absence of the Yang monopole. In analogy to the SO(5) invariant $S$, an SO(5) vector, defined as $X_a(x_a,n_i;x_a^{\prime},n_i^{\prime}) = \psi_{\alpha}(x_a,n_i) (\mathcal{R}\Gamma_a)_{\alpha\beta} \psi_{\beta}(x_a^{\prime},n_i^{\prime})$, transforms under the IRREP $(1,1)_{\text{\tiny {SO(5)} }}$ .
$X_a$ can be decomposed into an SO(4) scalar $X_5$ and  a 4-vector $X_{1\sim 4}$ transforming as $(\frac{1}{2};\frac{1}{2})_{\text{\tiny {SO(4)} }}$(see Table.~S5 S.M.~\cite{supplement}).
$S$ and $X_5$ is orthogonal to each other, since the former is an SO(5) invariant and the latter is a component of the five-vector. 
According to group theoretical analysis, the SO(5) of IRREP $(4I-2n+1,1)_{\text{\tiny {SO(5)} }}$ 
with $0\le n\le 2I$
can be decomposed into two branches of SO(4) IRREPs as 
$(4I-2n+1,1)_{\text{\tiny {SO(5)} }}  
= \bigoplus\limits_{j_1+j_2=2I-n} [(j_1,j_2)_{\text{\tiny {SO(4)} }} 
\oplus (j_1+\frac{1}{2},j_2+\frac{1}{2})_{\text{\tiny {SO(4)} }} ]$.
Hence, we have two sets of two-body states as 
\begin{widetext}
\begin{eqnarray}
&&\Phi_{J,M_1;K,M_2}^{(4I-2n+1,1)}(x_a,n_i;
x^\prime_a,n_i^{\prime}
) =  \mathcal{N}_{J,M_1;K,M_2}^{(4I-2n+1,1)} X_5 S^{n-1} f_{J,M_1,K,M_2}^{4I-2n} (x_a,n_i; x^\prime_a,n_i^{\prime}),  \nonumber\\
&&\Phi_{J+\frac{1}{2},M_1;K+\frac{1}{2},M_2}^{(4I-2n+1,1)}(x_a,n_i;
x^\prime_a,n_i^{\prime})= \mathcal{N}_{J+\frac{1}{2},M_1;K+\frac{1}{2},M_2}^{(4I-2n+1,1)}  
C_{\frac{1}{2},\alpha;J,m_1}^{J+\frac{1}{2},M_1} C_{\frac{1}{2},\beta;K,m_2}^{K+\frac{1}{2},M_2}   
			\Psi_{\frac{1}{2},\alpha;\frac{1}{2},\beta}^{(1,1)}
            S^{n-1} f_{J,m_1,K,m_2}^{4I-2n}(x_a,n_i;
x^\prime_a,n_i^{\prime})  
\label{eq:2-body2}, \ \ \
\end{eqnarray}
\end{widetext}
in which 
$\Psi_{\frac{1}{2},\pm\frac{1}{2};\frac{1}{2},\pm\frac{1}{2}}^{(1,1)}=\frac{1}{\sqrt 2}(X_1\pm i X_2) 
$, and
$\Psi_{\frac{1}{2},\pm\frac{1}{2};\frac{1}{2},\mp \frac{1}{2}}^{(1,1)}=
\frac{1}{\sqrt 2} (X_3\pm i X_4).
$ 
More generally, for $(n_1+n_2,n_1-n_2)_{\text{\tiny {SO(5)} }}$, the two-body states in these IRREPs are constructed from the higher-rank SO(5) harmonics, as detailed in Sec.~IV of
S.M.~\cite{supplement}.

To implement numerical calculations, we generalize the pseudo-potential formalism 
\cite{Haldane1983,Jainbook2007} from $S^2$ to $S^4$. 
The pseudo-potential, based on the two-body states constructed above, is represented by a series of projection operators as follows, 
\begin{eqnarray}
H=\sum \limits_{n_1,n_2} V_{n_1,n_2} 
P_{n_1,n_2},
\end{eqnarray}
where 
$P_{n_1,n_2}=\sum_\lambda 
|n_1, n_2,\lambda \rangle \langle n_1, n_2,\lambda|$
is the projection operator into the SO(5) IRREP $(n_1+n_2,n_1-n_2)_{\text{\tiny {SO(5)} }}$ and 
$\lambda$ denotes the SO(4) group quantum numbers associated with the state.
The interaction strength
$V_{n_1,n_2}$ is $\langle n_1,n_2 || V || n_1,n_2\rangle$, where $||$
means that the reduced matrix element only depends on the SO(5) IRREP.

We delve into the many-body microscopic wavefunctions. 
There are two different approaches to construct Laughlin wavefunctions, namely the determinant-type and
Jastrow-type the Laughlin wavefunctions, respectively \cite{Laughling1983,Haldane1983,zhang_2001_a}.
It is noteworthy that the determinant-type one (Eq.~(\ref{eq:determinant}) below) is not identical to the Jastrow-type one (Eq.~(\ref{eq:jastrow}) below). 
The difference has been highlighted in the study of the FQHE states on the $\mathbb{CP}^2$ manifold \cite{JieWang, DunghaiLee2006}.  
Nevertheless, both types of wavefunctions coincide on $S^2$.

\begin{figure}[t!]
\centering
\includegraphics[width=0.8\linewidth]{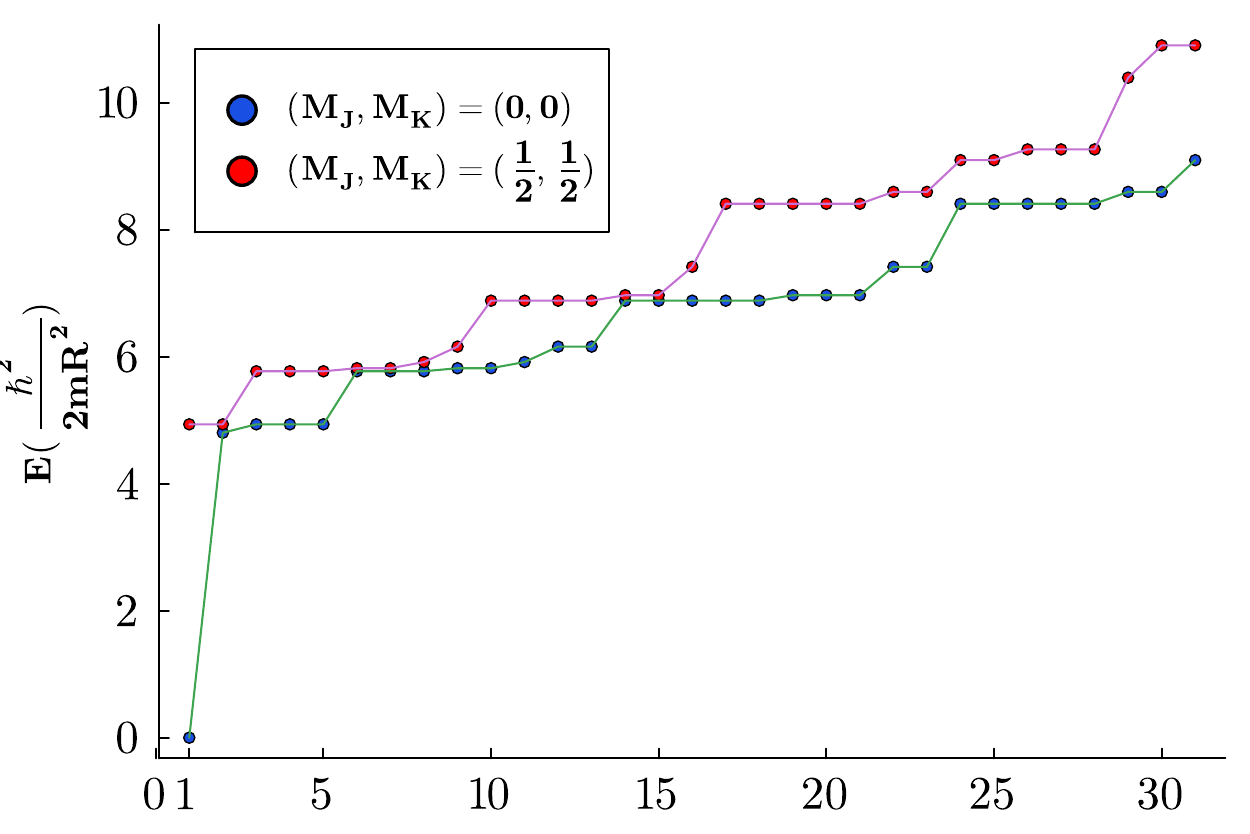}
\label{fig:main}
\caption{The energy spectra are shown in blocks marked by the quantum numbers $(M_J, M_K)$ for the Hamiltonian~(\ref{eq12}) with $N = 4$ particles occupying the LLL orbitals in the SO(5) IRREP $(3,0)_{\text{\tiny SO(5)}}$.
The two datasets
with
blue and red spots represent two branches of states with integer and half-integer SO(4) quantum numbers.
The number of orbitals  $N_o$ is 20.
The energy unit is taken as $\hbar^2/(2MR^2)$, and the horizontal axis shows the index of states 
in the ascending order in terms of energy. 
The zero energy ground state is an SO(5) singlet and unique. 
The lowest excitation is another SO(5) single state with a gap $\Delta=4.8$.
\label{fig:spectra} }
\end{figure}

The determinant-type Laughlin wavefunction takes the form of the Slater determinant raised to the $m$-th power, 
\begin{equation}
\begin{split}
& \Psi_{\text{det}}^{m} (x_{1,a},n_{1,i}; \ldots x_{N,a},n_{N,i} ) = \left[ \varepsilon_{P_1\ldots P_N} f_{A_1}^{(2I,0)}(x_{P_1,a},n_{P_1,i}) \vphantom{f_{A_N}^{(2I,0)}(x_{P_N,a},n_{P_N,i})} \right. \\
&\quad \left. \vphantom{\varepsilon_{P_1\ldots P_N} f_{A_1}^{(2I,0)}(x_{P_1,a},n_{P_1,i})} \times f_{A_2}^{(2I,0)}(x_{P_2,a},n_{P_2,i}) \cdots \times f_{A_N}^{(2I,0)}(x_{P_N,a},n_{P_N,i}) \right]^{m},
\end{split}
\label{eq:determinant}
\end{equation}
where $N$ is the particle number，
$A_h$ is the abbreviation of the SO(4) IRREP $(j_h,m_{h,1};k_h,m_{h,2})$
for the $h$-th particle, and $P$ is permutation 
of
1 to $N$. 
The filling of 
the state of
Eq.~(\ref{eq:determinant}) is $\nu=\frac{d(2I,0)}{d(2mI,0)}$, 
which approaches $1/m^3$ in the large-$m$ limit.  

We focus on the determinant Laughlin wavefunction with $m=3$ 
in the case of the smallest Hilbert space.
The 
wavefunction of a full-shell with $I=\frac{1}{2}$, i.e., $N=4$, is $\Psi_{\mbox{det}}=\varepsilon_{P_1P_2P_3P_4}
\Pi_{k=1}^4\psi_k (x_{P_k,a},n_{P_k,i})$,
and the corresponding determinant Laughlin state is
\begin{eqnarray}
\Psi^{(m=3)}_{\mbox{det},N=4}=
\left(\varepsilon_{P_1P_2P_3P_4} \prod\limits_{j=1}^{4} \psi_{j}(x_{P_j,a},n_{P_j,i}) 
\right)^3
\label{eq:N=4}
\end{eqnarray}
By counting the power of SO(5) spinors, the single-particle LLL orbitals involved in Eq.~(\ref{eq:N=4}) 
correspond to the case of $mI=\frac{3}{2}$ with the degeneracy $d(3,0)=20$.
The full-shell determinant state is an SO(5) invariant. Its power, the Laughlin determinant state, is likewise an SO(5) singlet. 
Upon exchanging two particles, the wavefunction acquires a phase factor $(-1)^3=-1$, 
ensuring the
Fermi statistics.
Hence, when expanding Eq.~(\ref{eq:N=4}), the relative wavefunction between a pair of particles $k$ and $l$ only contains
\begin{eqnarray}
\left(\mathbf{\Psi}(x_{k,a},n_{k,i}) \mathcal{M} \mathbf{\Psi}(x_{l,a},n_{l,i}) \right)^3,
\label{eq:contract}
\end{eqnarray}
where the anti-symmetry matrix kernel $\mathcal{M}$ takes either $\mathcal{R}$ or $\mathcal{R}\Gamma^a$.
Each anti-symmetric combination between two particles is denoted as one contraction.
For example, the structure of  Eq.~(\ref{eq:contract}) exhibits 3 contractions.


\begin{figure}[t!]
\centering \subfigure{\includegraphics[width=0.8\linewidth]{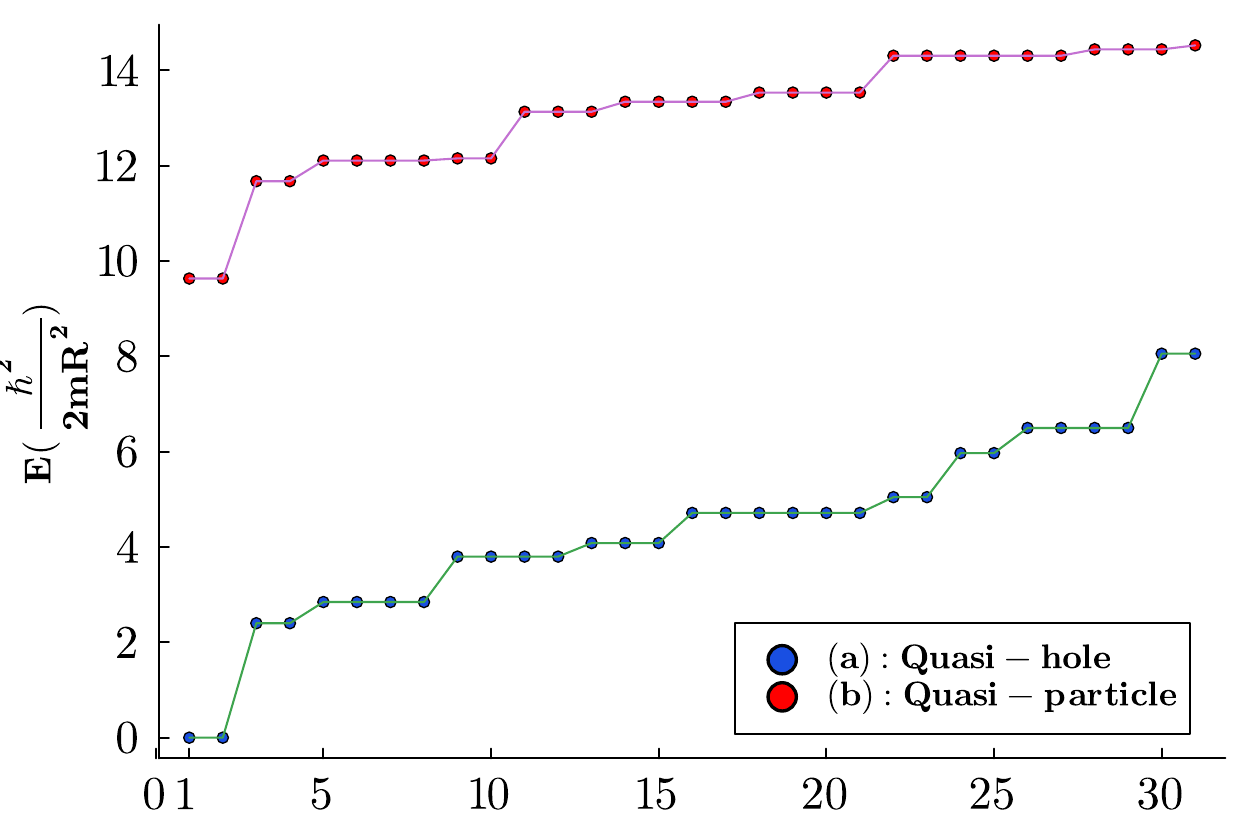}} 
\caption{The energy spectra of the same system as
in Fig. \ref{fig:spectra}:  ($a$) 
quasi-hole states with $N=3$,  ($b$) 
quasi-particle states with $N=5$.
Only the block of $(\frac{1}{2},0)$ is represented. 
The ground states of the quasi-hole case remain at zero energy, while those of the quasi-particle case exhibit a finite gap. 
The ground states for both the quasi-hole and quasi-particle cases belong to the 20-dimensional SO(5) IRREP \((3,0)_{\text{\tiny {SO(5)} }}\).
\label{fig:particle_hole}}
\end{figure}

Now we construct the pseudo-potential Hamiltonian for the many-body wavefunction
Eq.~(\ref{eq:determinant}).
The pseudo-potential Hamiltonian, constructed from two-body projection operators, is positive-definite and annihilates the wavefunction 
Eq.~(\ref{eq:determinant}).
Consequently, the annihilated wavefunction is a zero-energy ground state.
For concreteness, the special  case of Eq.~(\ref{eq:N=4}) with $m=3, N=4$ is employed as an example .
Consider two fermions in the LLL orbits with $I^\prime=mI=\frac{3}{2}$.   
The two-body states satisfying the fermionic statistics are decomposed into SO(5) IRREPs of $(4I^\prime-2r-1-s,2r+1-s)_{\text{\tiny {SO(5)} }}$ with $0\le r<I^\prime$ and $0\le s \le 2r+1$.
According to the two-body wavefunctions given in Eq.(VI.7) in  Sec.~VI of
S.M.~\cite{supplement} , the contraction number equals $2r+1$.
Hence, the IRREPs with $r=0$ , $(4,0)_{\text{\tiny {SO(5)} }}$ and $(5,1)_{\text{\tiny {SO(5)} }}$, are orthogonal to the two-body sectors of the wavefunction in Eq.~(\ref{eq:N=4}), as revealed by the contraction numbers.
Consequently, these two pseudopotential channels annihilate the wave function, which is the zero-energy ground state.
The above analysis can be easily generalize to an arbitary value of $N$
at $m=3$.
Then the pseudo-potential Hamiltonian is constructed as
\begin{align}
H& =\sum\limits_{1\le k< l\le N }\Big \{V_{(4mI-1,1)}  P_{4mI-1,1}(k,l)  \nonumber \\
& +V_{(4mI-2,0)}  P_{(4mI-2,0)}(k,l) \Big\},
\label{eq12}
\end{align}
where $P_{m,n}(k,l)$ is the projection operator of two-body states of the $k$-th and $l$-th 
particles.



Next we present the exact digonalization (ED) results for the case of $m=3$ and $N=4$. 
The SO(5) is a rank-2 Lie group with two commutable generators, which can be taken as the sum over all particles  of the diagonal operators of two inter-commutable SU(2) subgroups defined in Eq.~  (\ref{eq:so4_gen}) : $J_3=\sum_{l=1}^N J_3 (l)$ and $K_3=\sum_{l=1}^N K_3(l)$.
The pseudo-potential Hamiltonian is diagonalized in blocks labeled by different values of $(M_J, M_K)$, which are $J_3$ and $K_3$ eigenvalues, respectively.
Previously analysis shows that the wavefunction of Eq.~(\ref{eq:N=4}) is an SO(5) singlet and  ground state of the Hamiltonian~(\ref{eq12}).
Diagonalizations are performed and the ground state is found non-degenerate, hence, the wavefunction of Eq.~(\ref{eq:N=4}) is the unique ground state. 
Energy spectra in the blocks of $(0,0)$ and $(\frac{1}{2},\frac{1}{2})$ are shown in Fig.~\ref{fig:spectra}. 
The energies of the lowest and next lowest excited states are very close
at $4.80$ and $4.93$ in the unit of
$\frac{\hbar^2}{2MR^2}$.
The lowest excited state is also an SO(5) scalar  and the next lowest excited states are 35-fold degenerate forming the SO(5) IRREP of $(4, 0)_{\text{\tiny {SO(5)} }}$.
Such an SO(5) IRREP is decomposed into different SO(4) IRREPs carrying integer quantum numbers of $(2,0)_{\text{\tiny {SO(4)} }}\oplus 
(1,1)_{\text{\tiny {SO(4)} }}\oplus(0,2)_{\text{\tiny {SO(4)} }}$ with dimensions of $5$, $9$, and $5$, respectively, 
and those carrying half-integer quantum numbers of
$(\frac{3}{2},\frac{1}{2})_{\text{\tiny {SO(4)} }} \oplus
(\frac{1}{2},\frac{3}{2})_{\text{\tiny {SO(4)} }}$, both of dimension 8, as shown in  Table.~S9 of S.M.~\cite{supplement}.

We now present the spectra of excitations of the quasi-hole ($N=3$) and quasi-particle ($N = 5$) sectors as shown in Fig.~\ref{fig:particle_hole} ($a$) and ($b$), respectively.
Since each SO(5) IRREP contains SO(4) IRREPS
$(j_1,j_2)_{\text{\tiny {SO(4)} }}\oplus(j_2,j_1)_{\text{\tiny {SO(4)} }}$ symmetrically, we 
only present the state in the sector
$(\frac{1}{2},0)$ in terms of $(M_J, M_K)$
as a representative.
As for the quasi-hole states of $N=3$, its ground states remain at zero energy with the 20-fold degeneracy, lying in the SO(5) IRREP of $(3,0)_{\text{\tiny {SO(5)} }}$, which is decomposed to 
the SO(4) IRREPs of $(\frac{3}{2},0)_{\text{\tiny {SO(4)} }}\oplus (0, \frac{3}{2})_{\text{\tiny {SO(4)} }}\oplus (1,\frac{1}{2})_{\text{\tiny {SO(4)} }} \oplus (\frac{1}{2},1)_{\text{\tiny {SO(4)} }}$.
In contrast, as for the quasi-particle states
with $N = 5$, there exist a finite 
excitation gap.  
The ground states form the SO(5) IRREP
of $(3, 0)_{\text{\tiny {SO(5)} }}$, which is the same as the quasi-hole case.



\begin{figure}[t!]
\centering            \includegraphics[width=0.45\textwidth]{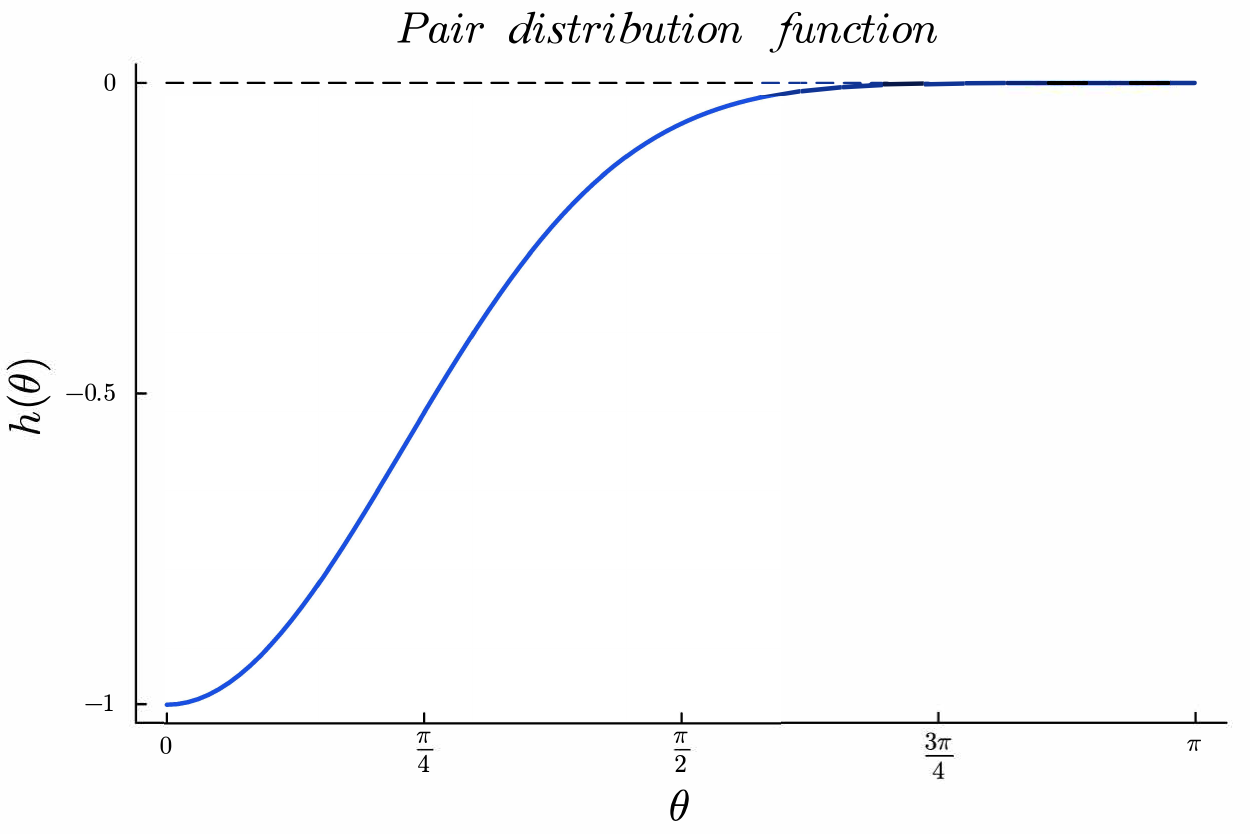}
\caption{The pair distribution function of the 4D determinant-type Laughlin wavefunction lacks a peak at long distance, indicating liquid-like behavior.
\label{fig:correlation} }
\end{figure}

Early studies of FQHE debated whether the ground state should be an incompressible quantum liquid or a crystalline  phase such as a Wigner crystal 
\cite{Anderson1979,Halperin1983a,Yoshioka1983,MacDonald1983,Yoshioka1984}.
The pair distribution function is frequently employed to reveal the incompressible nature of quantum liquids \cite{Jainbook2007}. 
We calculate the probability $h(\theta)$ of two particles separated by the chord distance, which depends only on $\theta$ due to the rotation symmetry. 
It is defined as 
\begin{align}
h(\theta) = \frac{\int d^3 n_1 d^3 n_2 \langle \Psi_0 \mid \rho(x_a,n_1) \rho(x_a^{\prime},n_2) \mid \Psi_0\rangle }{\rho(x_a)\rho(x^\prime_a)}
-1
\end{align}
where $|\Psi_0\rangle$ is the ground state of Hamiltonian~(\ref{eq12});  $\cos\theta 
=x_{a} x_{a}^{\prime} $;
$\rho(x_a,n_1)$ is the density operator;
$\rho(x_a)=\int d^3 n_1 
\langle \Psi_0| \rho(x_a,n_1)| \Psi_0\rangle $.
The calculated behavior of $h(\theta)$ is shown in Fig.~\ref{fig:correlation}, which is analogous to the case of the FQHE on the $S^2$ sphere~\cite{Haldane1985a,Girvin1986}.  
As $\theta$ goes to $0$,  $h(\theta)$ approaches $-1$ due to the Fermi statistics. 
Its amplitude features a smooth decay to zero as $\theta \rightarrow \pi$ without exhibiting pronounced peaks.
This result implies the absence of crystalline order, and is consistent with the quantum liquid like ground state in 4D.

        

Now we discuss the construction of the Jastrow-type Laughlin wavefunction on $S^4$ based on the SO(5) singlet 
$\psi_\alpha(x_1,n_1) \mathcal{R}_{\alpha\beta} \psi_\beta(x_2, n_2)$, which is a generalization of the SU(2) singlet function $u_iv_j-u_jv_i$ \cite{Haldane1983}.
The explicit wavefunction is expressed with the assistance of $\mathcal{R}$,
\begin{align}
\Psi_{\text{Jastrow}}^{m}
= \prod\limits_{1\le k <l \le N} \left(\psi_{\alpha}(x_{k,a},n_{k,i}) \mathcal{R}_{\alpha\beta} \psi_{\beta} (x_{l,a},n_{l,i}) \right)^{m},
\label{eq:jastrow}
\end{align}     
The power of the spinor $\psi$ of each particle appearing in 
Eq.~(\ref{eq:jastrow})
  is equal to $l=m(N-1)$, hence, the filling should be $\nu=N/ d(l,0)$ with $d(l,0)$ defined in Eq.~(\ref{eq:degneracy})   approaching $1/(m^3 N^2) \to 0$ as $N\to \infty$.
The parent Hamiltonian to this wavefunction is constrained in the channels of $\bigoplus\limits_{n_1+n_2 \geq l-m+1 }(n_1+n_2,n_1-n_2)_{\text{\tiny {SO(5)} }}$.
There is no well-defined thermodynamic limit for the Jastrow-type wavefunction. 
    
	
In conclusion, we have studied the 4D incompressible quantum liquid based on the Landau level under the SU(2) Yang monopole, specifically using the determinant-type  and Jastrow-type Laughlin wavefunctions. 
The parent Hamiltonian for these wavefunctions was constructed within the pseudo-potential formalism. 
The ED calculations show that the unique ground state  is an SO(5) singlet.
In addition, quasi-hole states remain at zero energy, while quasi-particle excitations are separated by a finite gap.
The liquid-like behavior is further supported by the pair distribution function.
However, a direct numerical analysis of both determinant-type Laughlin and Jastrow wavefunctions for larger particle numbers is prohibitive due to the enormous size of the Hilbert space. 
For the next accessible system size, 
$N$ equals 10 corresponding to the IRREP of $(2,0)_{\text{\tiny {SO(5)} }}$, and the LLL orbitals at $m=3$ correspond to the IRREP of $(6,0)_{\text{\tiny {SO(5)} }}$ which is 84 dimensional, then the Hilbert space dimension already exceeds  $10^{11}$. Nonetheless, the recent experimental demonstration of the few-electron Coulomb liquid \cite{Laughling2022nature,Laughling2024prl,PanJianwei-2024,Columbliquid2025} provides a promising platform for the experimental validation of our theoretical framework. 
Another intriguing avenue is to explore the statistical properties of excitations in the 4D fractional topological states, 
particularly investigating the loop or membrane statistics in higher dimensions~\cite{wu2010,WangLevin2014PRL080403,Cenkexu2023, Benevig2002,Zheng-cheng2019,YePeng2024,HuxfordNguyenKim2024MembraneOperators,Nair2025}. Furthermore, studying the entanglement spectra~\cite{Kitaeventropy,Xiaogang-entropy,Haldane2008}  offers valuable insights into these states and warrants further investigations.

 

\textit{Acknowledgment}.
--- We are grateful to Kun Yang, Hua Chen, Chen Lu,  L.Q. Chen,  Zhiming Pan, JianKeng Yuan, Jianyu Wang and Ziyuan Zeng  for stimulating discussions. 
C.W. is supported by the National Natural Science Foundation of China under the Grant Nos. 12550402, 
12234016, and 12574274.
W.Z. was  supported by National Science Foundation of China under Grants No. 12474144.	
This work has been supported by the New Cornerstone Science Foundation.

\bibliography{literature}

\newpage 
\section*{End Matter}

\textit{Yang monopole}--- 
Consider a quantum mechanical particle with an electric charge $e$ is constrained to $S^2$ enclosing a U(1) Dirac monopole with the monopole charge $g$.
The Dirac quantization condition leads to 
$eg/c=n\hbar/2$, where $n$ is equivalent 
to the first Chern number\cite{Dirac1931}.
It gives to wavefunctions described by two spinors: $u=\cos \frac{\theta}{2} e^{-\frac{\text{i}\phi}{2}}$ and $v=\cos \frac{\theta}{2} e^{\frac{\text{i}\phi}{2}}$.
When the $S^2$ sphere is replaced by $S^4$, 
the U(1) monopole should be augmented by the SU(2) one.
Correspondingly, the Dirac quantization 
condition is also generalized \cite{Yang197812}, which is equivalent to the second Chern class.
\begin{align}
   \mathcal{C}_2=\frac{1}{8\pi^2}\int_{S^4} \text{Tr} (F\wedge F) \label{second_chern}
\end{align}
Here, $\mathcal{C}_2$ is the 2nd Chern number, and $F=\frac{1}{2}\eta_{\mu\nu}^{a}I_{a} e^{\mu}\wedge e^{\nu}$, which in the vielbein is the curvature of the Yang monopole.
The SU(2) generators $I_a$ are in the spin-$I$ representation.
$\mathcal{C}_2$ can be evaluated as,
\begin{align}
    \mathcal{C}_2&=\frac{1}{32\pi^2}  \text{Tr}(I_a I_b)\int_{S^4} \eta_{\mu\nu}^{a} \eta_{\rho\sigma}^{b} e^{\mu}\wedge e^{\nu} \wedge e^{\rho}\wedge e^{\sigma}   \nonumber \\
    &=\frac{2I(2I+1)(2I+2)}{6} \label{second-chern}.
\end{align}
We use the identity of 't Hooft symbols at the third equality: $\eta_{\mu\nu}^{a} \eta_{\rho\sigma}^{a}=\delta_{\mu\rho}\delta_{\nu\sigma}-\delta_{\mu\sigma}\delta_{\nu\rho}+\varepsilon_{\mu\nu\rho\sigma}$ \cite{Hooft1976}. The volume of $S^4$ is defined as
$
\text{Vol}(S^4)=\int_{S^4} \frac{1}{4!} \varepsilon_{\mu\nu\rho\sigma} e^{\mu}\wedge e^{\nu} \wedge e^{\rho} \wedge e^{\sigma}.
$
Similar to the case on $S^2$, wavefunctions of 4D QHE require description by four Hopf spinor components, accounting for the half-spin carried by the SU(2) gauge field.

\textit{\text{so}\text{\small{(5)}} Lie Algebra} --- We define the mechanical angular momentum $\Lambda_{\mu\nu}=-\text{i} (x_{\mu} D_{\nu}-x_{\nu} D_{\mu})$, where $D_{\mu}=\partial_{\mu}-\text{i}A_{\mu}$. The curvature $F_{\mu\nu}$ is $F_{\mu\nu}=\text{i} [D_{\mu},D_{\nu}]$. Now examine the commutation relation of $\Lambda_{\mu\nu}$,
\begin{eqnarray}
&&[\Lambda_{\mu\nu},\Lambda_{\rho\sigma}]=\text{i} (\Lambda_{\mu\rho} \delta_{\nu\sigma}-\Lambda_{\mu \sigma}\delta_{\nu\rho}-\Lambda_{\nu\rho}\delta_{\mu\sigma}+\Lambda_{\nu\sigma}\delta_{\mu\rho})
\nonumber \\
&+&\text{i}(x_{\mu}x_{\rho} F_{\nu\sigma }-x_{\mu}x_{\sigma} F_{\nu\rho}
-
x_{\nu}x_{\rho}F_{\mu\sigma}+x_{\nu}x_{\sigma} F_{\mu\rho}).
\end{eqnarray}

The mechanical angular momentum does not satisfy the so\text{\small (5)} algebraic structure. We introduce the canonical angular momentum $L_{\mu\nu}=\Lambda_{\mu\nu}-R^2 F_{\mu\nu} $, which obeys the so\text{\small (5)} algebra
\begin{align}
[L_{\mu\nu},L_{\rho\sigma}]&=&\text{i} (L_{\mu\rho} \delta_{\nu\sigma}-L_{\mu \sigma}\delta_{\nu\rho}-L_{\nu\rho}\delta_{\mu\sigma}+L_{\nu\sigma}\delta_{\mu\rho}).
\end{align}

We note that $\sum\limits_{\mu<\nu} \Lambda_{\mu\nu}^2=R^2\pi^2-(\vec{R}\cdot \vec{\pi})^2+3\text{i}(\vec{R}\cdot \vec{\pi})$, where $\pi_{\mu}=-\text{i} D_{\mu}$. 
The particle motion is restricted to the sphere $S^4$ of radius $R$, so $\vec{R}\cdot \vec{\pi}=0$. Hence, the Hamiltonian is expressed in terms of $L_{\mu\nu}$,
\begin{eqnarray}
   H&=&\frac{1}{2mR^2}\vec{\pi}^2=\frac{1}{2mR^2}\sum\limits_{\mu<\nu} \Lambda_{\mu\nu}^2\nonumber \\
   &=&\frac{1}{2mR^2}\sum\limits_{\mu<\nu} (L_{\mu\nu}^2-F_{\mu\nu}^2).
   \label{SO5-Landau}
\end{eqnarray}

For the SU(2) gauge field components defined as $F_{\mu\nu}=\eta_{\mu\nu}^{a}I_a$, the sum of the squared field strength yields $\sum_{\mu<\nu}F_{\mu\nu}^2=2I(I+1)$. 
This result follows from the contraction $\eta_{\mu\nu}^{a}\eta_{\mu\nu}^{b}=4\delta^{ab}$. The eigenvalues of $\sum_{\mu<\nu} L_{\mu\nu}^2$ are given by the SO(5) Casimir $\mathcal{C}(N+2I,N)$ \cite{Yang1978}.

\begin{table}[h!]
\centering
\footnotesize
\renewcommand{\arraystretch}{1.6}
\setlength{\tabcolsep}{3pt}
\caption{Comparison between $S^2$ and $S^4$ Quantum Hall Systems}
\label{tab:comparison_narrow}
\begin{tabularx}{\columnwidth}{@{} l >{\raggedright\arraybackslash}X >{\raggedright\arraybackslash}X @{}}
\hline
Manifold & $S^2$ & $S^4$ \\
Gauge Field & U(1) & SU(2) \\
Quantization & $\displaystyle\frac{1}{2\pi}\int_{S^2} F \in\mathbb{Z}$ & $\displaystyle\frac{1}{8\pi^2}\int_{S^4} \text{Tr}(F\wedge F) \in\mathbb{Z}$ \\
Symmetry & SO(3) & SO(5) \\
Generator & $\vec{L} = \vec{r}\times\vec{\pi} - \frac{eg}{c}\hat{r}$ & $L_{\mu\nu} = \Lambda_{\mu\nu} - R^2 F_{\mu\nu}$ \\
Algebra & $[\hat{L}_i,\hat{L}_j] = \text{i}\varepsilon_{ijk}\hat{L}_k$ & 
$\begin{aligned}
&[L_{\mu\nu},L_{\rho\sigma}]=\text{i} (L_{\mu\rho} \delta_{\nu\sigma}\\ &
-L_{\mu \sigma}\delta_{\nu\rho}-L_{\nu\rho}\delta_{\mu\sigma}
\\ &
+L_{\nu\sigma}\delta_{\mu\rho})
\end{aligned}$ \\
\\
Hamiltonian & $\displaystyle H=\frac{\vec{L}^2-(\frac{eg}{c})^2}{2m R^2}$ & $\displaystyle H=\frac{\sum\limits_{\mu<\nu} (L_{\mu\nu}^2 - F_{\mu\nu}^2)}{2mR^2}$ \\
Fundamental spinor & 1st Hopf spinor \cite{Haldane1983} & 2nd Hopf spinor \cite{zhang_2001_a,supplement} \\
\hline
\end{tabularx}
\end{table}

\textit{Experimental aspect} --- 
Topological phases and the associated integer quantum Hall effects in four dimensions have been realized in various platforms, including cold atom systems, optical lattices, acoustic lattices, and electric circuits \cite{science2024, lohse_2018_exploring,zilberberg_2018_photonic,GuancongMaprx2021,Chen2021,Xu2022,Chen2023NSR,Chenyanfeng2024,Price2020,BaiLeZhang2020,Yu2020NSR4DSpinless,Zhang2020PRB4DHexadecapole}.
The internal degrees of freedom of atoms, typically the hyperfine spin components, are often employed as synthetic dimensions to simulate high dimensional systems, though the lengths of synthetic dimensions are finite \cite{science2024, lohse_2018_exploring}.  
Another method is via the 2D Thouless pumping, where a 2D system modulated by two additional adiabatic parameters $(\phi_x, \phi_y)$ is topologically equivalent to a 4D integer topological system \cite{Qi2008}. 

Strong repulsive interactions are necessary to achieve fractional topological states. 
In ultra-cold atom systems, Feshbach resonances are typically employed to tune inter-atomic interactions \cite{simulator}. 
Significant progress has been achieved in realizing interaction effects in Thouless pumping processes \cite{Walter2023HubbardThoulessPump,Zhou2023,Thouless-interaction-2024,Zhu2024}.
Alternatively, the plasmonium photon box array \cite{PanJianwei-2024} offers a compelling photonic platform, in which preliminary fractional quantum Hall signatures have been observed.
The key mechanism is the blockade effect derived from the substantial anharmonicity of plasmonium modes \cite{Jianwei-Plasmon}. 
This effect prohibits double occupancy on a single site, thereby naturally generating the strong repulsive interactions essential for fractional quantum Hall states. 


Hence, it is desired to integrating synthetic dimensions with interaction control (e.g., Feshbach resonances) to experimentally explore fractional topological states in four dimensions. 
In principle, Feshbach resonances could be designed between a series of carefully chosen pairs of internal states, for example, between every two neighboring sites along synthetic dimensions. 
In this case, the interaction would be short-ranged not only in 3-dimensional real space but also in synthetic dimensions with internal states. 

Certainly, designing feasible proposals to engineer strong interactions in desired forms to realize 4-dimensional fractional topological states would be highly non-trivial, which is certainly beyond the scope of this manuscript.
It would naturally be a potential focus for ultra-cold atom studies to bridge the gap between high-dimensional fractional topological states and realistic experimental platforms.
Our work servers a starting point to motivate and stimulate studies both experimental and theoretical. 
While the current study focuses on characterizing the ground state, future investigations into the dynamical properties and transport signatures of 4D quantum incompressible fluids are also desired. 
We hope our findings serve as a timely guidance to these ongoing efforts in quantum simulations.

\end{document}